\documentclass{article}

\usepackage{PRIMEarxiv}

\usepackage[utf8]{inputenc} 
\usepackage[T1]{fontenc}    
\usepackage{hyperref}       
\usepackage{url}            
\usepackage{booktabs}       
\usepackage{amsfonts}       
\usepackage{nicefrac}       
\usepackage{microtype}      
\usepackage{lipsum}
\usepackage{fancyhdr}       
\usepackage{graphicx}       
\graphicspath{{media/}}     
\usepackage{orcidlink}
\usepackage{subcaption}%
\usepackage{rotating}%
\usepackage{longtable}%
\usepackage{tabularray}

\title{Collaboration in Virtual Reality: Survey and Perspectives
\thanks{\textit{\underline{Citation}}: 
\textbf{Authors. Title. Pages.... DOI:000000/11111.}} 
}

\author{
  Ourania Koutzampasopoulou Xanthidou 
  \orcidlink{0000-0002-2659-7138} \\
  Computer Science, \\
  Brunel University London, \\
  Uxbridge, London, UB8 3PH, United Kingdom\\
  \texttt{Rania.KoutzampasopoulouXanthidou@brunel.ac.uk} \\
  \And
  Nadine Aburumman\orcidlink{0000-0003-4578-8738} \\
  Computer Science, \\
  Brunel University London, \\
  Uxbridge, London, UB8 3PH, United Kingdom\\
  \texttt{Nadine.Aburumman@brunel.ac.uk}
  \And
  Hanêne Ben-Abdallah\orcidlink{0000-0001-9215-4661} \\
  CIS \& Applied Media, \\
  Higher Colleges of Technology, \\
  Academic City, Dubai, United Arab Emirates\\
  \texttt{hbenabdallah@hct.ac.ae} \\
}

\begin{document}
\maketitle

\begin{abstract}
The application of Virtual Reality Environments (VRE) has been gaining momentum as a relatively new tool to assist with mitigating various difficulties including abstractness of concepts, lack of user engagement, perception of disconnection from other users. A VRE may offer both synchronous and asynchronous experiences, in addition to an immersive environment which promotes users’ engagement. Past research has shown that, in general, VRE do improve the experiences they try to enhance in many aspects of human activity. Terms like immersiveness and 3D representation of real life objects and environments are, as it appears, the two most obvious positive effects of Virtual Reality (VR) applications. However, despite these benefits it does not come without challenges. The main three concepts/challenges are the spatial design, the collaboration interaction between its members and the VRE, and the audio and video fidelity. Each of the three includes a number of other components that should be addressed for the total experience to be fine-tuned. These include mutual embodiment and shared perspectives, teleportation, gestural interaction, symmetric and asymmetric collaboration, physical and virtual co-location, inventory, and time and spatial synchronization. This paper comprises a survey of the literature, that identifies and explains the features introduced and the challenges involved with the VREs, and furthermore provides various interesting future research directions.
\end{abstract}


\section{Introduction}

It goes without saying that together with Data Analytics, and AI/ Machine Learning/ Deep Learning, extended reality (XR) is a term that almost everyone in the information technology field is talking all the more. VR, together with Augmented Reality (AR) and Mixed Reality (MR) are the three forms that constitute the umbrella of technological revolution referred to as XR. Although these three forms have many common features, they also have different challenges that need to be addressed, in particular in collaborative environments \cite{gronbaek2023partially}.

The history of VR (initially), and subsequently AR and XR, goes as far back as at least 25 years. However, in particular during the last decade, serious private investments have triggered rapid developments in the field \cite{cipresso2018past}. According to Statista, the revenue in the XR market is expected to reach US\$31.12 bn in 2023 and will keep growing by an annual average rate of 13.72\% until 2027 reaching a projected US\$52.05 bn \cite{statistaWorldwideStatista}. This, by itself, is a sufficient justification of the relative “hype” in the discussions about the value of XR in modern business and other activity worldwide. Kolmar (2023) reports that 23\% of the VR/AR device users are aged between 25 and 34, with 57\% of the device owners being male and 43\% of them female. The same study showed that 55\% of the respondents identified the cost of the devices as the top barrier for VR adoption. However, as the cost keeps dropping, one can only expect the adoption of XR will increase \cite{kolmar}. 

There is no industry that VR technology has not become prevalent in: Gaming \cite{meldrum2012virtual, zyda2005visual}, military \cite{alexander2017virtual}, architecture \cite{song2017improvement}, education \cite{englund2017teaching}, health care \cite{O'Connor2019virtual}, and government \cite{rea2022VRGovernment} are only a few of the fields VR/AR/XR has penetrated recently. The revenue from related products sold in these sectors has increased on average around six times between 2017 and 2022 in leading markets like China, Australia, Argentina, Brazil, Canada, Egypt, France, Germany, India, Indonesia, Italy, Japan, Mexico, Poland, Turkey, United Kingdom, United States, South Korea, and Spain just to mention some \cite{statistaWorldwideStatista}. 

It appears that one of the applications of VR is as a training tool for the employees. Indeed, although relevant research, long ago, has been focused on children’s education, the results point towards enhanced narrative-based, immersive, and a constructionist user experience through the application of VRE \cite{roussosnice}. 

There are three major issues associated with the traditional training process. First, there are many complaints about negative experiences due to poor training and low sense of belonging that might demotivate the trainees \cite{giannakos2017identifying}. A lot of work has been done towards increasing the level of motivation during the training process, often achieved by means of various gamification approaches perhaps through online learning tools that aim at motivating the trainees \cite{kenwright2016holistic, philips2015immersive, wang2021authenticity}. As to the quality of training, it is mostly addressed by enforcing the development of better teaching practices and strategies of the faculties and motivate the students through a virtual enhancing experience \cite{wee2022iprogvr}. To this end, quite often, trainers are at least encouraged to attend programs that lead them to teaching certifications, e.g., Microsoft’s Certified Educator, Fellow of Higher Education Academy, and similar \cite{rob2014certification}. 

Second, there is a sense of abstractness or even misunderstanding of many concepts in various fields especially those related to technology \cite{medeiros2018systematic}. Again, applying gamification during the training process has shown significantly positive results. 

Third, often the problem is lack of availability to attend training in a traditional classroom setting for a variety of reasons including the physical distance from the venue \cite{goodwinOnlineClasses}. The usual way to address this problem is by means of training through the use of online learning management system and/or video conference tools.

It is surprising, though, that although numerous studies have been conducted on the effect of the application of VRE in the above industries (as mentioned above), it is still difficult to find studies on the effect of collaborative VRE in such activities \cite{sasinka2018collaborative}. 
This is even though research studies, like \cite{yassien2021give} which focused on the effect of gender in a VR collaborative experience, have shown that collaborators with different demographics appear to have different levels of satisfaction and immersiveness of a VR experience \cite{vasilescu2015gender}. The problem is that in this and other similar studies the main research question is not the effect of collaboration but of the participants' demographics when collaborating. In all these cases, the role that collaborative VR has to play in enhancing relevant experiences could and should be further studied. 

In this state-of-art paper a survey of the relevant literature is presented and described concerning those shortcomings, challenges, barriers, issues faced during the implementation and deployment of VR solutions.

The rest of the paper is divided into three main sections which are briefly explained in the next on VRE as a Tool. First, the concept of Spatial Design with its aspects/attributes of mutual embodiment and teleportation is presented in detail. Second, the main focus on this paper, the concept of collaboration and interactivity and its aspects/attributes of gestural interaction, symmetric/asymmetric collaboration, physical co-location, and key points are discussed. Third, technical requirements related to visual and audio fidelity including inventory, time and spatial synchronization, and software, hardware and connectivity, are all explained. Finally, the paper concludes by suggesting a number of possible research questions that could be further studied as future research work.  All these literature, in particular the reviewed published experiments between 2011 and 2023, are briefly outlined in the Appendix as a table of the survey of experiments on the VRE (see \ref{Appendix}).

\section{VRE as a Tool}\label{sec2} 

VREs provide the facilities to mitigate the problem of explaining abstract concepts and ideas in the various fields that, as it appears, do not have a direct representation in the physical world \cite{singh2021, smutny2019}. They increase students’/trainees’ motivation by offering enhanced opportunities for gamification in the educational/training process by allowing to create virtual worlds to bring a greater sense of enjoyment to the learning process \cite{vanDer2016serious, wang2018critical}. Indeed, deBack et al. (2020), investigated the effect of gamification in the learning process by experimenting with 40 participant, between-subjects, on their CAVE VRE, i.e., their version of a VR classroom. They found that such VREs result in higher learning gains compared to traditional textbook learning  \cite{deBack2020benefits}. They offer the opportunity for learning/training away from traditional classrooms/venues and inside immersive environments, with tactile interactions, sound, and other senses’ perceptions that make the process both more interactive and more engaging \cite{akbulut2018effectiveness, sasinka2018collaborative, singh2021}. This is particularly useful when physical attendance is not possible at the venue but is required. In this case, the individual has the alternative to appear in the VRE as an avatar, i.e., in a form of a virtual representation of the user. 
This allows collaboration inside the environment, i.e., the concept of having two or more students/trainees/users working and interacting, regardless of their physical location \cite{branovic2013development, nersesian2020interdisciplinary}. VRE, as a revolutionary technology, are meant to accommodate such a requirement. This appears to be applicable in several situations in a variety of sectors (as mentioned in the introduction), with variable intensity and benefits \cite{cipresso2018past, sakkas2022applied}. 

The most important concepts of VRE are the following: 

\begin{itemize}
\item \textbf{Spatial Design:} This is a broad term used to describe a number of attributes of a VRE that make it more immersive. Several technical terms, including \emph{shared perspectives through mutual embodiment} \cite{zaman2015nroom} and \emph{teleportation} \cite{xia2018spacetime} are often used to describe the various types of experiences that may enhance immersiveness of the environment and ensure reliable communication between users \cite{zaman2015nroom}.
\item \textbf{Collaboration/Interactivity:} The complicated concept of having two or more users of the same VRE interacting with each other and with objects of the environment raising issues like co-location collision, take-over control, gestural interactions, and locomotion \cite{drey2022towards, moulec2016take}.
\item \textbf{Visual and audio fidelity:} An umbrella term used to address technical issues such as synchronization of animation and sound, visual info and communication, inventory, user experience visual fidelity synchrony, and more \cite{liu2018no}.
\end{itemize}
\noindent

\section{Spatial Design}\label{sec3}

Spatial design is a general term that refers to the general perception of the user(s) of a VRE about mutual embodiment and teleportation, discussed in detail in the next sections. In both cases the term "co-presence" is commonly used to refer to the feeling of the user of being actively involved with others and fully immersed as a personality in a collaborative VRE \cite{freiwald2021effects}. 

It can be measured by means of performance metrics inside the VRE, e.g., requesting the users to complete some tasks, or by asking the users to fill in a survey at the end of their experience to evaluate the validity, stability, and feeling of comfort of the environment. In both cases, either through measuring actual users’ performance or by means of a survey, what is measured is the users’ spatial or communication perception. The next sections describe, in brief, the above aspects/attributes of spatial design and co-presence.

\subsection{Mutual embodiment}\label{subsec3.1}
Mutual embodiment refers to the concept of synchronizing users’ actions and sharing attention of the same objects from the same vantage point, often also called shared perspective. A study by Zaman et al. (2015) suggested that this characteristic increases the reliability of the communication between users. Indeed, it showed that the efficacy of a collaborative VRE relies on the ability of the users to develop mutually compatible shared perspectives that should be interchangeable with other users in order to motivate them to engage with their actions and share a common understanding of the general topology of the VRE. The added feature of allowing verbal communication between the users in the VRE increased their level of satisfaction as they could mutually convey their embodied understanding of the spatial qualities of the VRE \cite{zaman2015nroom}.

One of the main challenges implementing such a mutual embodiment VRE is to overcome the spatial ambiguities caused by the difference in the positions and orientations of the users \cite{zaman2015nroom}. In other words, it is important that the 3D objects are designed with careful details to make their attributes not just visible to all users of the VRE but also in the same details and without spatial inconsistencies depending on the user’s position and/or orientation. This is a rather serious challenge and very important to address if the goal is the visual accuracy of the environment as a whole and the reliability of the 3D objects’ visibility and other attributes. 

Hoppe et al. (2021), developed a ShiSha prototype which featured a shifted share approach (hence ShiSha), in which the user's avatar has the position in the VRE shifted to the side to avoid collision with other users' avatar, when visualizing 3D object in a VRE and compared it with the Vishnu approach \cite{leChenechal2016vishnu} in which no such shift occurs and the users' avatars may co-exist in the same location in the VRE. The ShiSha experiment involved 32 participants and found that the ShiSha approach had higher usability than the Vishnu approach, higher task performance and higher co-presence \cite{hoppe2021shisha}.

\begin{figure}[ht!]
\center
  \includegraphics[scale=0.5]{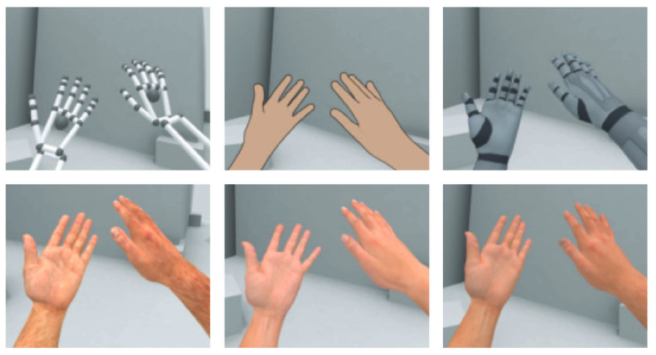}
  \caption{\centering Styles of virtual hands \cite{Schwind2017-vq}}
  \label{fig.1:Styles of virtual hands}
\end{figure}

Another challenge is to achieve high level of details of the user’s avatar’s \cite{freiwald2021effects} and hands \cite{Schwind2017-vq} appearance as studies showed they increase the level of embodiment hence raising the immersiveness and interaction quality as well as the overall user experience. In the case of hands, Schwind et al. (2017) investigated the effect of different designs and styles of hands, i.e., human and non-human, on the preferences of the users based on their gender. During the experiment of 28 students and employees of a university with an average age of 26.07 years, the users had to perform three different tasks to ensure the hands are present in the field of view of the participant and facilitate an immersive VR experience. A VR questionnaire appeared in front of them within the VRE. The results showed that females using male hands felt very uncomfortable but males using female hands found them very realistic and unusual but attractive. Also, deviations from the movement of participants’ own bodies (i.e., hands) were felt as uncomfortable and were negatively criticized regardless of gender. Finally, females felt higher levels of co-presence using non-human hands in contrast to males \cite{Schwind2017-vq}. Fig. \ref{fig.1:Styles of virtual hands} illustrates six different styles of virtual hands. In the top row, the styles are all non-human (abstract, cartoon and robot, from left to right respectively). In the bottom row, the styles are all human (male, female, and androgynous, from left to right respectively).

A particular decision has to be made in the case of the use of virtual hands as mechanisms for gestural interactions. Hrimech et al. (2011) referred to these 3D mechanisms as interaction metaphors. They examined the impact of these mechanisms, in their experiment where they invited 32 participants in pairs, of ages between 18 and 57, to collaborate in a 3D puzzle game in three conditions. The users could see only the interaction metaphors representing their partner without any representation of avatars. The three interaction metaphors proposed are \cite{hrimech20113d}: 
\begin{enumerate}
    \item The Ray Casting metaphor with which the user can select an object through a virtual ray pointing it.
    \item The virtual hand metaphor which appears as a normal human-virtual hand and functions as one.
    \item The GoGo metaphor, an "elastic" virtual arm activated automatically when the user wants to interact with distant objects.
\end{enumerate}

It must be noted that Hrimech's et al. (2011) study revealed a feeling of disappointment when using virtual hand metaphor because the users were expecting finger animation when interacting with objects and this simply was not available in their experiment. On the contrary, the use of the Ray Casting metaphor showed that the users had higher levels of involvement compared to the other two. Finally, the GoGo metaphor gave higher levels of collaboration effort compared to the other two \cite{hrimech20113d}.

\subsection{Teleportation}\label{subsec3.2}
Teleportation is a locomotion technique for navigating in a VRE by pointing and pushing/releasing the VR pointing device or touch controller to a particular object or location in the environment, often called an intersect, resulting in instantly moving to the particular location of the object in the VRE \cite{thanyadit2020substituting}.

It is an effective technique, especially in those cases where the VRE is considerably large resulting in time and movement inefficiencies when navigating in it. There are at least four identified challenges to address when implementing the teleportation mechanism, i.e., teleportation technical requirements, visualizations, scaling of objects, and group teleportation.

Concerning the teleportation's technical requirements, the whole process must have four quality attributes. Firstly, it must be time efficient, meaning its duration should be independent of the teleportation distance otherwise it would reduce the benefit of teleportation which is to save time to move to a particular location in the VRE. Second, it must be traceable providing the teleporter’s origin and destination and the path taken during the teleportation process, so that the other users may trace it. This can be achieved using a stack structure that holds the log of teleportation history of the user as in Xia et al. (2018) experiment \cite{xia2018spacetime}. Third, it must be intuitive with the users being able to instinctively, at a glance and based on previous experience with other visual human computer interfaces, understand how to proceed with the process. Finally, it should be easily recognizable so that it is easy to interpret the user’s intention prior to the move \cite{thanyadit2020substituting, weissker2020getting}.

Thanyadit et al. (2020), suggested four different methods that could be followed. The first is the "hover" method with which "the avatar moves in a straight line toward the destination at a constant speed while turning toward the target destination". The method resembles the assumed motion of walking which makes the process traceable, intuitive, and recognizable. There are two notable drawbacks of the method. One is that it is not time efficient as it takes time to move to the intersect point, but it could be bypassed by just accelerating the movement process to the destination. The other is the significant likelihood to collide with objects in the way \cite{thanyadit2020substituting}.

\begin{figure}[ht!]
\center
  \includegraphics[scale=1]{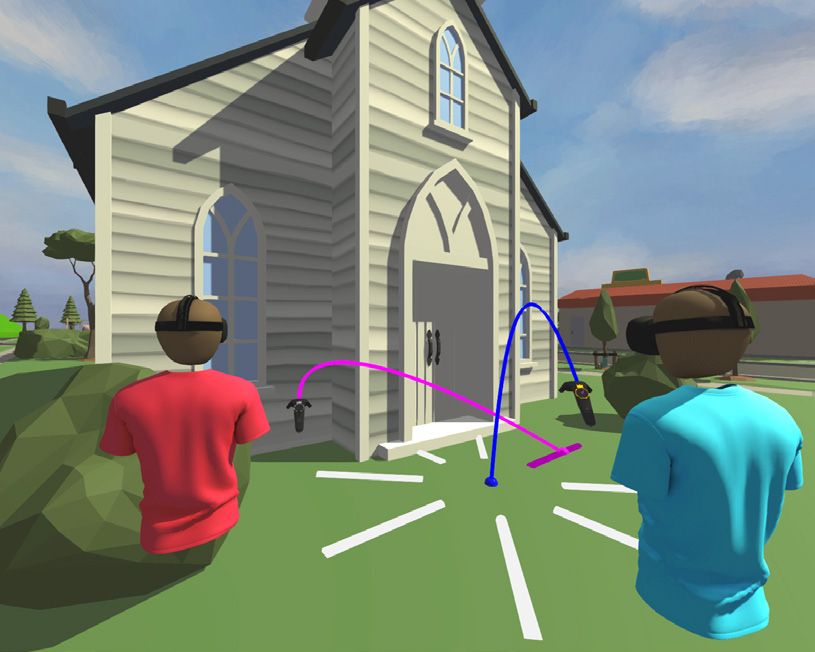}
  \caption{\centering The Jumping technique for teleportation: \textit {"Our two-user jumping technique for remote collaboration allows the navigator (blue) to adjust the translational offset of the
passenger (red) when planning a jump."} \cite{weissker2020getting}}
  \label{fig.2:JumpingTechnique}
\end{figure}
 
The second method is the "jump" method (Fig. \ref{fig.2:JumpingTechnique}) that can be used as the answer to the second problem of the "hover" method explained before. Following this method, the avatar appears to be moving in a parabolic trajectory towards the destination like as in an actual jumping motion. This, obviously and largely, solves the problem of collision with objects in the way \cite{thanyadit2020substituting, weissker2020getting}.

A third method could be simply, the "fade" method which suggest fading the avatar from the original location to the destination. It is "recognizable and time-efficient, but not traceable nor intuitive" \cite{Schwind2017-vq, thanyadit2020substituting}. Finally, a fourth method is the "portal" which suggests two portals with the avatar fading from the one at the origin before appearing at the exit portal. The method is traceable, recognizable, and time efficient as it is almost instant, however, it might be not quite intuitive depending on the user’s experience with VRE \cite{kato2022reality}.

Thanyadit et al. (2020) concluded from their study that the "hover" method is the preferred choice. It would be interesting to study a combination of the "hover" with each of the other three methods in order to define what is the optimal solution to use. It would also be interesting to study which of the above four techniques/methods is more appealing for the user of the VR experience \cite{thanyadit2020substituting}.

The third challenge when implementing teleportation locomotion is to manage the scaling of the size of the users’ avatars depending on their location in the VRE in relation to the other users. This is important because it enhances the feeling of spatial awareness of the VRE and the correct sense of distances amongst its users.  Another challenge is the decision whether the other users should be somehow informed, perhaps through a greyed avatar, of one’s teleportation from a particular location to another in the VRE and, if so, to which one \cite{xia2018spacetime}.

The fourth challenge relates to the technical problems that are raised when two or more VR users are navigating the same environment in symmetric mode and with the same goals while moving and/or interacting with the environment. In such cases it is often necessary to group the users to have joint navigation. Weissker et al, (2020) explored the possibilities of navigation techniques in a VRE, experimenting with 40 participants in pairs, to derive a framework in which they suggested a four-step process to address this technical issue, i.e., forming groups, ``norming'' by assigning responsibilities between the members of the groups, performing through groups, and adjourning the groups. They also found that allowing users to switch between individual and group navigation can be beneficial for the collaborative work of specially distributed participants. Additionally, it appeared that group navigation helped the users to stay together to focus on the joint observation, discussion, and evaluation of virtual content. Finally, they concluded that the addition of virtual formation adjustments allowed navigators to resolve problematic situations arising during group jumping and to direct passenger attention to interesting features without the need to giving verbal navigation instructions. \cite{weissker2020getting}.

Forming groups can be achieved through various ways. According to Weissker et al., (2020), it can be done automatically by the system as a ``circumstantial'' action based on the proximity of the users or predefined spatial arrangements. It can be done automatically when certain ``environmental'' conditions are meant, e.g., two or more users happen to enter in a particular object, say, a vehicle. It can be explicitly created by a single user through a singular or a mutual confirmation process but consequently selected and confirmed by the various users of the environment \cite{weissker2020getting}.

Assigning responsibilities to group members, i.e., ``norming'', can be established in four different ways. It can be that all members can equally input their travel preferences and the system is tasked to combine them to an optimum one \cite{lessel2019hedgewarssgc}. It can be that the members contribute their input but with different weights as to the overall system decision \cite{lessel2017crowdchess}. A third option is to have the travel controls restricted to a single ``navigator'' with the rest just following as ``passengers'' \cite{weissker2019multi}. Finally, it is possible to allow the system to auto-generate travel inputs based on pre-defined paths in which case all VR users are treated as ``passengers'' \cite{galyean1995guided}. Galyean (1995) suggests that, in the latter two cases, the mechanism might be following a confirmatory process, in which the users are required to confirm the choice, or contradictory, in which they only react if they reject the offer.

Performing the tasks as groups is considered by Weissker et al. (2020) as the single most important stage of this process. The members must be able to communicate with each other, must be fully aware of both the VRE in detail and the other members details and must be well-informed of the groups’ goal. Specific details must be given in that regard \cite{weissker2020getting}.

At the end of the process, there must be a mechanism to stop the group which is referred to Weissker et al., (2020) as adjourning the groups \cite{weissker2020getting}.

Finally, an added point of concern is whether the users indeed find teleportation the best approach or not. Ardal et al. (2019), suggest that it is, often, more intuitive to navigate in a VRE by moving the virtual space and scaling it up or down instead of teleporting \cite{ardal2019collaborative}. Another study suggested that teleportation (which implies non-continuous movement) causes more cybersickness symptoms than the usual and is often not preferred as less enjoyable \cite{freiwald2021effects}. The VRE designers/developers should not take the benefits of teleportation for granted but, instead, explore the various possibilities. 

\section{Collaboration and Interactivity}\label{sec4}

Wang et al. (2021), define interaction as "the level of responsiveness the VRE provides to the user" \cite{wang2021authenticity}. The term refers to several aspects related to the behaviour of a user of a VRE including gestural interaction, non-verbal interactivity, locomotion, and take-over control, to name some of the most important. 

As to collaboration, as early as back in 1998 Churchill and Snowdon defined \emph{Collaborative Virtual Environments (CVEs)} as "shared spaces designed to support interactions between users and objects in Virtual Reality" \cite{churchill1998collaborative}. Hrimech et al. (2011) identified three key components that must be addressed when designing the experience \cite{hrimech20113d}:
\begin{enumerate}
    \item The actions of the partner users,
    \item The intentions of the partner users, and
    \item The partner users’ perspectives of the environment.
\end{enumerate}

In collaboration, “positive interdependence” refers to the users’ ability to complete tasks as a team. Whether various experts in a field collaborate in the VRE to solve a complex problem, or students in a class try together as groups to understand the concepts at hand and work on projects, having clear rules and roles with clear responsibilities among the users helps improve effectiveness of addressing the tasks \cite{sasinka2018collaborative}. In a similar fashion, like in the previous case of spatial design and co-presence, either one or both alternatives, i.e., measured performance or surveyed evaluation, can be used to measure the validity and efficacy of the interactions in the environment, usually referred to as interaction perception.

In a successful CVE, the components of both collaboration and interaction must be seamlessly coupled. The sections that follow describe in some detail the above components, the possible benefits and whatever shortcomings related to them.

\subsection{Gestural Interaction}\label{subsec4.1}

Gestural interaction is the ability to place action on various objects of the VRE or other users in it. The more detailed is the design of the object the better the interaction experience \cite{sasinka2018collaborative}. Mine (1995) organized conceptually the interaction process into four distinct and equally important components, i.e., navigation, selection and validation, manipulation, and system control \cite{mine1995isaac}. 

By navigation, one refers to the cognitive process of developing a map of the VRE where the user is expected to move in \cite{laviola20173d} and the “motor” component by which the user either moves in the environment or is transported from location to location \cite{hrimech20113d}. 

When the user is ready, and has decided which object to interact with, the next step is to proceed with the selection of the object by pointing to it and validating it. The former is done usually by means of a “ray”-like technique with which the user points to the object depending on its distance from the user. This is accomplished by a variety of methods one of which could be to display, on the spot, the identity of the object together, perhaps, with some of its core attributes and possible interactions with it \cite{hrimech20113d, mine1995isaac}. 

\begin{figure}[ht!]
\center
  \includegraphics[scale=0.9]{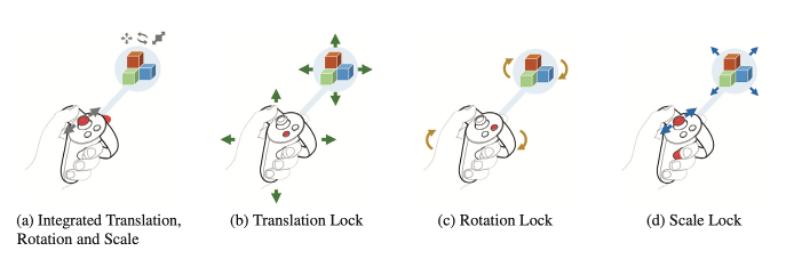}
  \caption{\centering Interaction gestures for distant manipulations with one hand. \cite{grandi2019characterizing}} 
  \label{fig.3:Interaction gestures}
\end{figure}

\begin{figure}[ht!]
\center
  \includegraphics[scale=0.75]{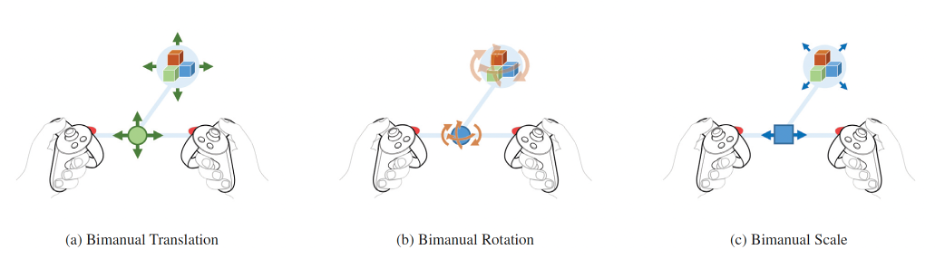}
  \caption{\centering Bimanual transformations \cite{grandi2019characterizing}.}
  \label{fig.4:Bimanual transformation}
\end{figure}

Once the object to interact with is selected, then the next step is to translate it, i.e., to manipulate it by modifying such properties of the object like its size, texture, or transparency, rotate it, i.e., to change its possible orientation and position, and scale it, i.e., to change its size in the different dimensions \cite{hrimech20113d},. Grandi et al. (2019), developed a 3D docking task that required aligning the position, rotation, and scale of the controlled virtual object with an identical target object. In Fig. \ref{fig.3:Interaction gestures} (b) the user changes the shape of the object, in (c) the orientation by rotating the object, and in (d) the scale of the object, i.e., its size. All these the user does with a single hand. The integration of all these manipulation in a single move is illustrated in (a). Fig. \ref{fig.4:Bimanual transformation} illustrates the same manipulations of a single user using both hands. The same apply if more than one users collaborate in an interaction with a 3D object \cite{mapes1995two, grandi2019characterizing}.  

An experiment by Zaman (2015) suggested that a lack of haptic feedback affects in a negative way the user from placing action on objects or making gestural interactions to other users. Actually, the study claims that there is an indication this lack of haptic causes users to even misinterpret the distance of 3D objects in the environment thus failing to correctly interact with those \cite{zaman2015nroom}. 

During gestural interaction, that is the object’s manipulation, there are a few decisions to make. One is to determine the characteristics of the haptic feedback the user gets when interacting with the 3D objects, i.e., the characteristics of possible vibration during the interaction \cite{kataoka2019interactiveHapticDevice}. This can be measured by means of the duration of the vibration, its frequency, and its amplitude \cite{choi2017grabity}. A number of research studies conducted recently addressed the effect of haptic feedback in the form of vibration in the user’s experience of the VRE \cite{xia2018spacetime, zaman2015nroom, webb2022haptic, wang2020haptic}. The second decision is the form, type, and orientation of the sound originating from the objects and towards the user. 

The third decision is how to control the 3D objects that the user attempts to interact with when they are in distance. In order to address this, the concept and mechanism of the extended arm may be implemented but carefully so that it does not distort the whole environment \cite{fidalgo2023magic}.

\begin{figure}[ht!]
\center
  \includegraphics[scale=0.45]{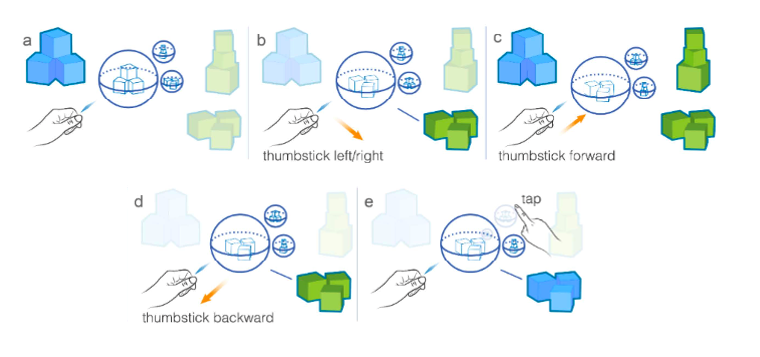}
  \caption{\centering "Manipulating Parallel Containers. a) Parallel Containers, b) Switching parallel version, c) Highlighting all the versions, d) Hiding all the versions except the chosen one, e) Merging a parallel version into the chosen one" \cite{xia2018spacetime}.}
  \label{fig.5: Manipulated Parallel Container}
\end{figure}

Finally, the fourth decision is how to have an object manipulated by multiple users in a collaborative VRE simultaneously. The same object may be available in more than one location and state at the same time for the various users to interact or, instead, it can only appear in one location and state with the users changing those potentially real time. The former case, i.e., the object appears in several locations and states, one for each user of the VRE, was addressed in an experiment by Xia et al. (2018) with six professionals, of ages between 22 to 41, through their 3D editing tool called SPACETIME in which two to three could collaborate and edit a scene of 3D objects simultaneously by working on copies of the same object. Each copy of a user is represented as in a sphere in which the version of the object appears. This is called a container as it appears in Fig. \ref{fig.5: Manipulated Parallel Container} (a). If the same or another user wants to manipulate another copy of the object, the user only needs to select a different container as in Fig. \ref{fig.5: Manipulated Parallel Container} (b). It is also possible to highlight all containers just to select the one interested to manipulate Fig. \ref{fig.5: Manipulated Parallel Container} (c). After selecting a container, the user can hide the other containers Fig. \ref{fig.5: Manipulated Parallel Container} (d). In this case, it will not be too long before the objects in the containers become different as to their locations and states, as a result of the users’ interaction with them, causing the whole environment for the different users to not reflect the same reality for all. They referred to this scenario as parallel objects in which eventually the manipulated objects collapse in one version in which the consistent/same states are automatically merged and those inconsistent/not same are pending for approval by the various users \cite{xia2018spacetime}. This last step can be seen in Fig. \ref{fig.5: Manipulated Parallel Container} (e) in which the final version of the manipulated object is decided and the rest are just deleted.  

In the latter case, i.e., the object appears in one location and state for all the VR users, the new decision is to control which user is the most realistic case to have the priority to be allowed to interact with the object \cite{xia2018spacetime} and, perhaps, what should be the appearance of the object during that instance. The other users interested to make changes to the object must wait their turn on the copies/instances of the object. Every time a change is made on an object, automatically the other instances of the object are updated. The implementation of this technique may vary and it might be similar as the containers of Xia et al. (2018) or it might be completely different. Sutherland (1964) refers to this technique as instancing of 3D objects \cite{sutherland1964sketch}. 

The integration of the above three functionalities/points, i.e., navigate, select and manipulate objects, with the VRE is fundamental and is the fourth component of this process. It is referred to as system control and is the mechanism that makes the options and functions of the VRE, including its various 3D objects, available to the user when connected \cite{hrimech20113d, mine1995isaac}.

Another term quite relevant to the above is the one referred to as take-over control. It is the concept of having two or more users of a VRE interacting over the same object(s) and, potentially, changing its status, i.e., position, form, etc. Moulec et al. (2016) attempted to describe in detail the main elements of take-over control and their relationship through a maritime navigation system, tested by 8 participants in pairs, within-subjects, in which a trainee was tasked to avoid an under-water obstacle with the help of a trainer. The obstacles could be in the form of rocks or coral and not visible. The trainee had the control of the boat but the trainer could assume control at any time if the boat was about to collide with an obstacle. The study showed that the users preferred smooth transition over the sudden one by the trainer and experienced an increased level of awareness and informativeness when the visual feedback was present of the obstacles \cite{moulec2016take}.

\subsection{Collaboration: Symmetric v. Asymmetric}\label{subsec4.2}

Collaboration is the concept of having two or more users of a VRE placing workflows in parallel \cite{fleury2010generic}. Ideally, a collaborative VRE should facilitate that its users can not only discuss, edit, validate, etc., different tasks and navigate through it, but also create new scenes and tasks in it, as was the case of Ardal et al. (2019) and their experiments \cite{ardal2019collaborative}. 

\begin{figure}[ht!]
     \centering
     \begin{subfigure}[b]{0.45\textwidth}
         \centering
         \includegraphics[width=\textwidth]{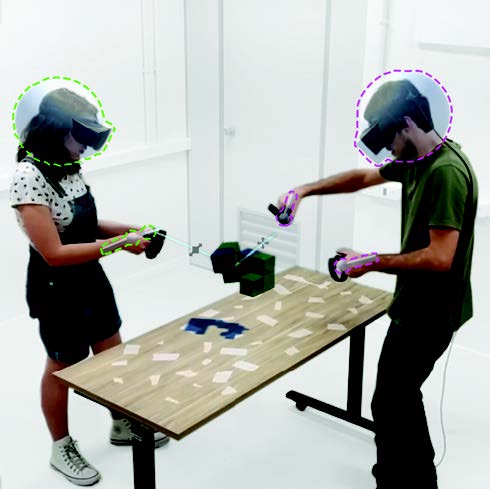}
         \label{Symmetric VR}
     \end{subfigure}
     \hfill
     \begin{subfigure}[b]{0.45\textwidth}
         \centering
         \includegraphics[width=\textwidth]{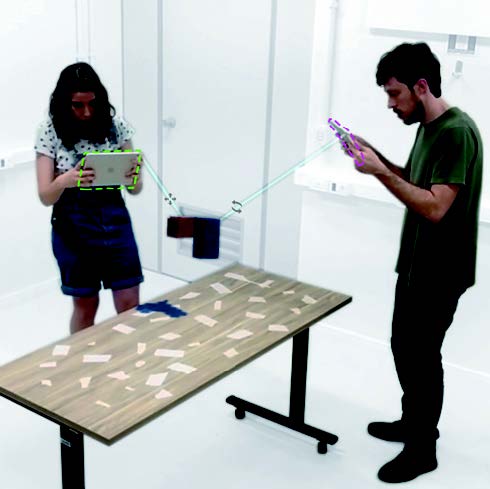}
         \label{Symmetric AR}
     \end{subfigure}
     \hfill
        \caption{Left image: Symmetric VR where both users use VR headset. Right image: Symmetric AR where both users use tablet/mobile \cite{grandi2019characterizing}.}
        \label{fig.6:Symmetric VR and AR}
\end{figure}

\begin{figure}[ht!]
\center
  \includegraphics[scale=0.9]{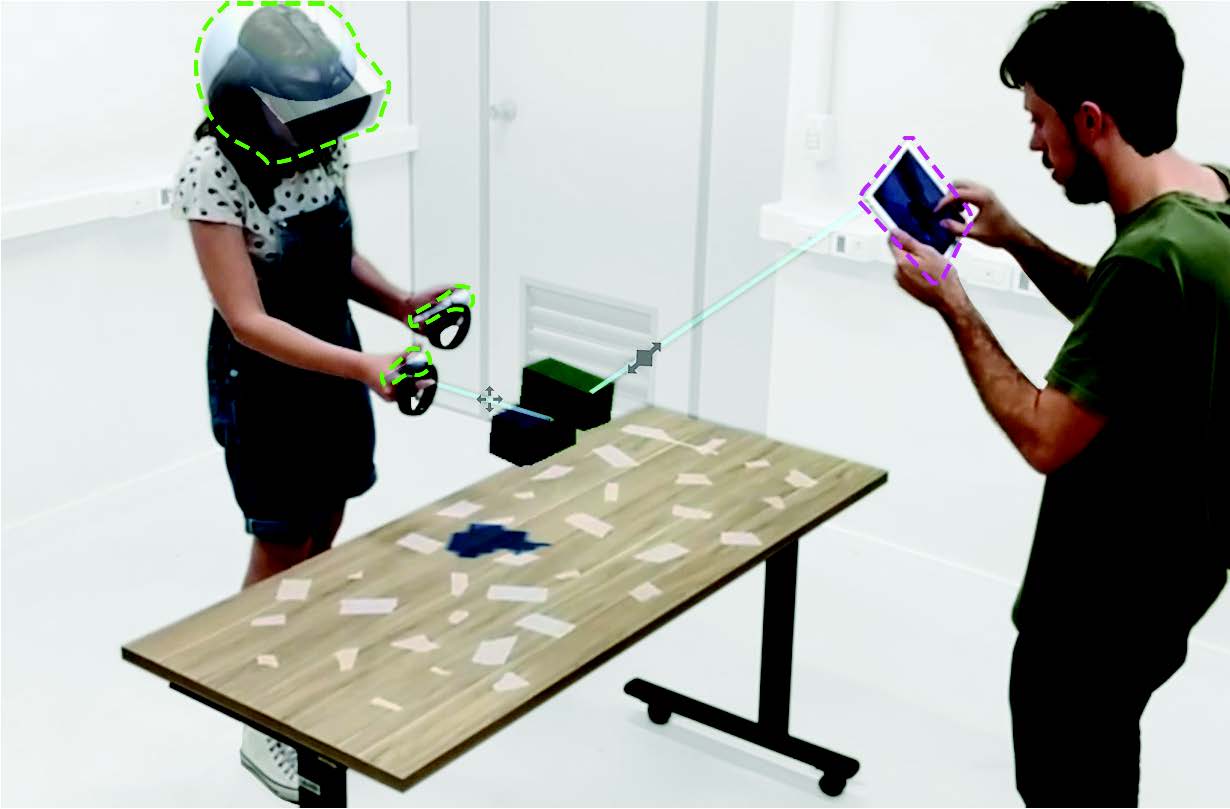}
  \caption{\centering Asymmetric VR and AR where one user uses VR headset and the other uses tablet or mobile \cite{grandi2019characterizing}.} 
  \label{fig.7:Asymmetric VR/AR}
\end{figure}

Collaboration in VRE may take two forms, i.e., symmetric and asymmetric, depending mainly on how the 3D content is perceived by the users and the type of interaction that may be placed on it, if at all. In the case of symmetric VR collaboration, all users collaborate in the same VRE wearing VR headsets or, again all, use their tablets or mobiles (Fig. \ref{fig.6:Symmetric VR and AR}). In the case of asymmetric VR collaboration, some users wear VR headset while others use their tablet or mobile (Fig. \ref{fig.7:Asymmetric VR/AR}) \cite{drey2022towards, grandi2019characterizing}. 

Grandi et al. (2019), developed a 3D docking task that required aligning the position, rotation, and scale of the controlled virtual object with an identical target object. The idea was to recreate the physical room in a VRE and have two users to collaborate in it. They tested their VRE with 36 participants in pairs (within-subjects) and found that although working in pairs did not lead to an increase in speed during the task resolution, perhaps because of the simplicity of the task, it reduced significantly the workload. Additionally, they found that VR-VR condition outperforms both the AR-AR and VR-AR conditions \cite{grandi2019characterizing}. 

In the case of symmetric, also referred to as “egocentric” \cite{weissker2020getting, zaman2015nroom}, all users are within the VRE experiencing the same 3D objects and, potentially, interacting directly on them in parallel workflows as equals. Drey et al. (2022) in their study, compared symmetric with asymmetric collaboration testing their VR/AR prototype with 46 conveniently sampled participants, in a within-subjects experiment. They found that both students and teachers preferred the symmetric collaboration as been more helpful and fun, improving the learning process. They also found that teachers experienced difficulties in orientation in the VRE when following the asymmetric mode and had to rely on the students, following the symmetric mode, to ensure correct understanding of the environment. This caused communication problems between students and teachers in the symmetric/asymmetric mode \cite{drey2022towards}. Hence, it appears that symmetric collaboration helps the users recall objects in the VRE more accurately and, therefore, there are indications it might improve the learning process as compared to the traditional learning environments by enhancing the motivation towards the tasks at hand and increasing the sense of co-presence, communication and immersiveness.

The very nature of symmetric collaboration, though, raises major technical issues that must be addressed and, indeed, efforts are already made in the recent past or are currently under way. One such issue is the possibility of having one of the users obstructing the view of a 3D object or the interaction with it from the others, often referred to as occlusion \cite{fidalgo2023magic}. Another concern is whether it is preferred to have the VR users collaborating face-to-face or side-by-side. In the case of the former, i.e., face-to-face, the possibility of occlusion is increased, and the users do not share the same viewpoint of the 3D objects given their opposite orientation. However, it might be more beneficial as it appears it enhances the level of communication between the users. In the case of the latter, Fidalgo et al. (2023) tested their system called MAGIC following a within-subjects experiment of 12 conveniently sampled participants to find out that there is a higher pointing agreement for both participants when sharing perspectives, i.e., in a side-by-side collaboration, yielding a better understanding of the collaboration tasks. Following this approach,  the users share the same viewpoint of the 3D object, making the whole experience more object-centred, and dramatically reduce, perhaps even eliminate, the likelihood of occlusion. However, it should be noted that Fidalgo's study did not find any significant difference between the time to complete the experiment when following the case of side-by-side as compared to the face-to-face collaboration \cite{fidalgo2023magic}. Yet a third concern is what happens in the case of more than 2 users collaborating symmetrically in the VRE in which case, obviously, the face-to-face and side-by-side approaches don’t apply directly, and more research need to be conducted on this.

The symmetric (egocentric) type of collaboration appears to be more suitable for small-scale manipulation of objects that allows for better precision when the objects are within the user’s reach \cite{hrimech20113d, zaman2015nroom}.

In the case of asymmetric collaboration (also referred to as exocentric collaboration) \cite{weissker2020getting, zaman2015nroom}, usually there are pairs users; the explorer wears the Head-Mounted Display (HMD) and navigates inside the VRE and the navigator moves in the same environment from outside through a touchscreen tablet or any other form of a monitor and a pointing device through what can be seen as a bird’s-eye view. In the case of educational VRE, it is often the case, the explorer is the learner, and the navigator is the teacher \cite{drey2022towards}. 

In asymmetric collaboration, some of the concerns and disadvantages of symmetric collaboration are, by default, not present. There is no concern of face-to-face or side-by-side arrangement of the VR users, as the second of the two is only “supervising” the tasks and guiding the first user without, usually, having direct interaction with the 3D objects and the environment. For the same reason, there is no issue of possible occlusion or other types of conflicts between the users and neither a possibility of having distortion of the position and the states/orientation of the 3D objects by the two users.

There are, however, in this case of asymmetric collaboration, different challenges that must be addressed. One is the level of detail of the objects and setting of the VRE provided to the explorer and to the navigator, i.e., whether it needs to be the same or not. Another is the type of exact roles and resources available to each of them. Yet a third one is to address the problem that, apparently, the two different types of users will have clearly different experiences from the same environment and, hence, evaluate the experience differently \cite{wang2021authenticity}.

Finally, in both symmetric (egocentric) and asymmetric (exocentric) collaboration, it is important to decide if it is better for the users of the VRE to belong to the same knowledge domain or, instead, if it is better to have an interdisciplinary team of users \cite{nersesian2020interdisciplinary}.

\subsection{Physical Co-location}\label{subsec4.3}

In the case of symmetric (egocentric) collaboration, this might take, again, two different forms, namely, co-located and distant collaboration. Co-located collaboration is the term that refers to the concept of having two or more VR users in the same physical room and rather close to each other. Distant collaboration suggests that the users are at different locations which implies that, by default, there is no possibility of physical collision between the users \cite{rios2018users}.

Ideally, in the case of co-located collaboration, each user's avatar representation and position in the VRE must match exactly the user's physical location. However, if the VRE is significantly larger than the physical room, where the users are physically located, this may cause spatial desynchronization. This is because as a user moves in the VRE the other users of the same environment will be able to see this user’s location in the VRE but not the actual physical location. Obviously, this possibility raises safety issues resulting from possible collisions of the physical bodies of the users, especially when the users are wearing HMDs and walk in a cohabitation workspace, i.e., in the same physical room \cite{lacoche2017collaborators}. In turn, this also causes reduction of velocity speed and considerable caution of movement. 

\begin{figure}[ht!]
\center
  \includegraphics[scale=0.36]{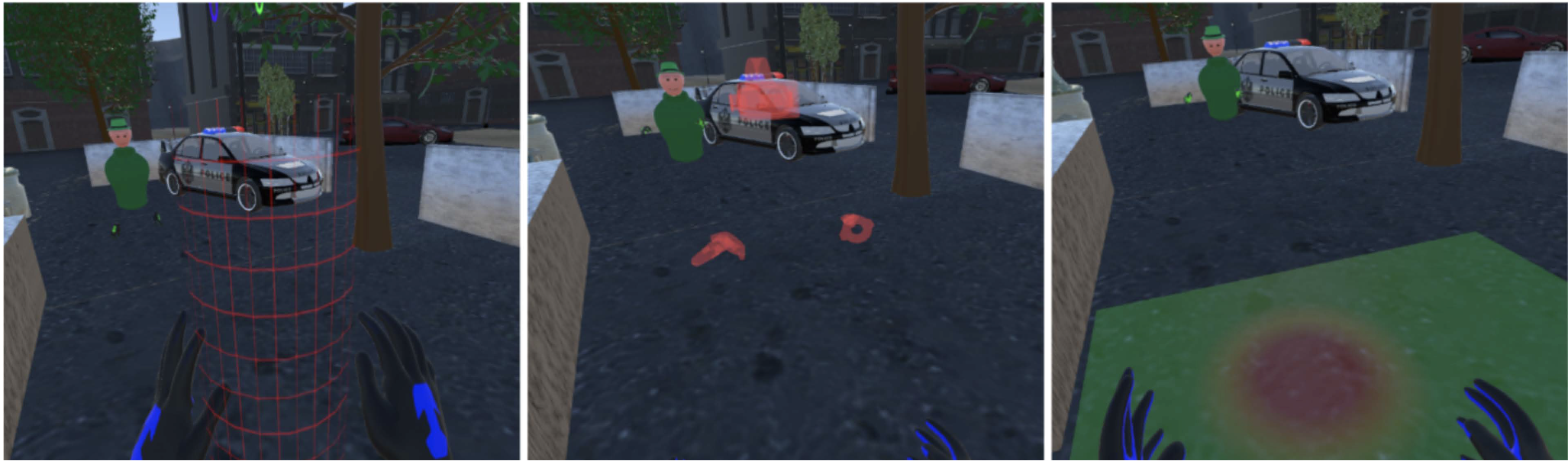}
  \caption{\centering (a) The "Extended Grid" (left), (b) The "Ghost Avatar" (middle), and (c) The "Safe Navigation Floor" (right) metaphors \cite{lacoche2017collaborators}.}
  \label{fig.8:The Co-Location metaphors, EG,GA, SNF}
\end{figure}

Lacoche et al. (2017) examined this problem of users' cohabitation and safety in the shared physical workspace when they can navigate both virtually and physically. They ran a shooter game experiment with 38 participants in pairs, of ages between 14 and 59, most without previous VR experience but most with 3D gaming experience. Each pair played a different version with different conditions, i.e., in Extended Grid (EG), Ghost Avatar (GA), Navigation Floor (SNF), and Separated Tracked Spaces (STS). The first technique/metaphor is the EG in which there is a grid in a cylindrical shape that appears around another user as the main user comes closer to the other. The second, the GA, makes the avatar of the other user appear as a ghost (in a variety of ways) with only the controllers or the virtual head and hands appear. The third technique, the SNF, is the concept of allowing the users to navigate in a safe zone when the ground is green coloured or not allowing if the ground is red coloured. In the latter case, the users are always visible to all others, and they don’t obstruct the field view of the main user. Of all the four, GA is considered as the safest, with less physical collision, followed by EG. These conditions are illustrated in Fig. \ref{fig.8:The Co-Location metaphors, EG,GA, SNF}. In contrast to what was hypothesized, the study did not prove that the safety of the participants is better with the STS compared to the other conditions. This is probably because the users are forced to constantly look at the floor instead of the game and each other. Also, it was not verified that the users felt more free in their physical movements under the EG, GA, and SNF conditions rather than under STS condition \cite{lacoche2017collaborators}. 

\begin{figure}[ht!]
\center
  \includegraphics[scale=0.5]{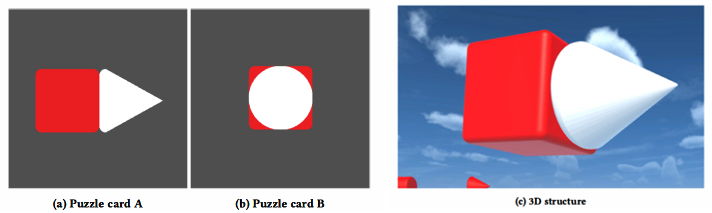}
  \caption{\centering CoBlok puzzle \cite{wikstrom2022coblok}.}
  \label{fig.9:The CoBlok puzzle}
\end{figure}

Rios et al. (2018) studied the user's locomotor behavior during an experiment in which 8 participants in pairs, of ages between 20 and 40, shared the same virtual space of a physical indoor lab room 2x4 meters in which they had to collaborate to perform a specific task. The task involved them walking from one to the other side of the VRE, grab the pieces of a puzzle and place them in the correct order. They could view the VRE and could hear the foot steps through earphones. They found out that the synchronization or not of animation and footsteps sound did not affect the participants' performance. Also, it appeared that the participants in immersive VREs were more cautious than in the real world, no matter how accurate and realistic the VRE looked, and no matter how small the anomalies in the synchronization of animation and sound, as they were worried about colliding in the physical room while wearing their HMD  \cite{rios2018users}. However, a relevant experiment by Sanchez and Slater (2005) indicated that the general behavior of the participants in a VRE collaborating with others is rather similar to that in their real life even though their performance may change \cite{sanchez2005presence}. In any case, there is a need to measure the performance of collaboration in physical co-location environments to make relevant useful conclusions like with the case of the CoBlok application. Wikstrom et al. (2022) introduced and evaluated the CoBlok application, a puzzle in which participants in pairs received cards of 3D shapes from different 2D views and they needed to decide the correct 3D version. Fig. \ref{fig.9:The CoBlok puzzle} (a) and (b) illustrate the 2D views of the 3D shapes from different cards for each user in the pair. The users must be able to decide quickly and accurately what the correct 3D depiction is (c), without any tools in their hands but only through verbal communication and without been able to see each other's card \cite{wikstrom2022coblok}. 

Liu and Kaplan (2018) examined the two coordination processes affected by the loss of visual information in collaborative VRE, i.e., situation awareness and conversational grounding. In their experiment, with 16 participants in pairs, of ages between 20 and 35, all with prior computer gaming experience, they found that physically co-located test subjects tend to move and talk less when independently solving puzzles inside an isolated VRE. Furthermore, when other players’ avatars and biometrics are introduced, then, the players get more stressed and competitive. Additionally, the ability to work on a problem in parallel (VR split) fosters more communication than taking turns, and also leads to less overall test-subject movement. Finally, keeping collaborators working independently leads to lower sense of urgency than turn-based collaboration \cite{liu2018no}.

Many of the aforementioned concerns are, by default, absent in the case of distant (also known as remote) collaboration. For example, the likelihood of physical collision of the users is eliminated for obvious reasons. This, in turn, secures more spatial consistency and higher environment fidelity \cite{fidalgo2023magic}. Another one could be whether combining two techniques, e.g., EG and GA reduces the likelihood for co-location collisions. A third could be to explore the possibility of collisions when users move backwards \cite{lacoche2017collaborators}.

Another serious challenge of collaboration is when the users’ avatars share the same position in the VRE. This problem is often referred to as overlapping avatars or avatar collisions and raises visual fidelity issues \cite{hoppe2021shisha, weissker2020getting}.  Weissker et al. (2020) suggest that it can be achieve by means of fixing the positions of the avatars with minimum and maximum placement distances between them \cite{weissker2020getting}.

\subsection{Key Points}\label{subsec4.4}

There are a few key points one must address before preparing a VR experience. To start with, one should not take for granted that a collaborative VRE is the preferred choice for every different task and scenario. Ardal et al. (2019) ran their experiment with 20 professional filmmakers, of ages between 18 and 59, in within-subjects mode, in which they found their collaborative environment was more enjoyable and marginally more preferred than solo experiment. Also, the avatars and audio feedback during the collaborative mode of their experiment made the communication aspect between the users successful. This may be because when the users are represented as avatars and verbal communication is allowed, then, they are able to discuss and decide immediate actions. This enhances the feeling/sense of co-presence which, for some, is quite important and a determinant factor for the satisfaction of their experience \cite{ardal2019collaborative}. 

However, the professionals performance showed statistically significantly better results under the solo condition. This, they suggested, depends on the personality of each user as, e.g., extroverted individuals may prefer collaborative environments as opposed to introverted individuals, who may feel better alone. In the case of the latter, and given the value of collaboration in general, if the solo setup is performed first, then, the collaborative setup might be easier and, hence, more attractive even for these types of people \cite{ardal2019collaborative}.

Another suggestion by Ardal et al. (2019) is that more attention should be given on the design of the VRE in as much detail as possible and less on the collaboration and communication of its users. For example, it is important to pay more attention to the details of the representation of the users faces, body, etc. as well as facial expressions, gestural interactions, body movements and dialogues, than to the specifics and detailed aspects of how collaboration will take place \cite{ardal2019collaborative}. This is particularly useful, perhaps even critical, in those cases where there are more than just two users collaborating in the VRE \cite{liu2023synthesizing}.

Finally, Ardal et al. (2019) suggest that specialized roles and common tasks and functions for particular VR users, especially in the case of exocentric collaboration, is key to make the experience more effective \cite{ardal2019collaborative}.

\section{Visual and Audio Fidelity}\label{sec5}

Since a VRE makes intense use of visual and audio features and memory, one can only expect visual and audio fidelity (i.e., quality) to be among the most important aspects of such an environment, especially as quite often such environments are set up to run on an organization’s cloud that adds the extra complication of taking care of internet bandwidth required to run such environments. There are some key visual and audio features that would enhance the user’s experience in a VRE and a few of the most important are described in the following sections.
 
\subsection{Inventory}\label{subsec5.1}

\begin{figure}[ht!]
\center
  \includegraphics[scale=0.35]{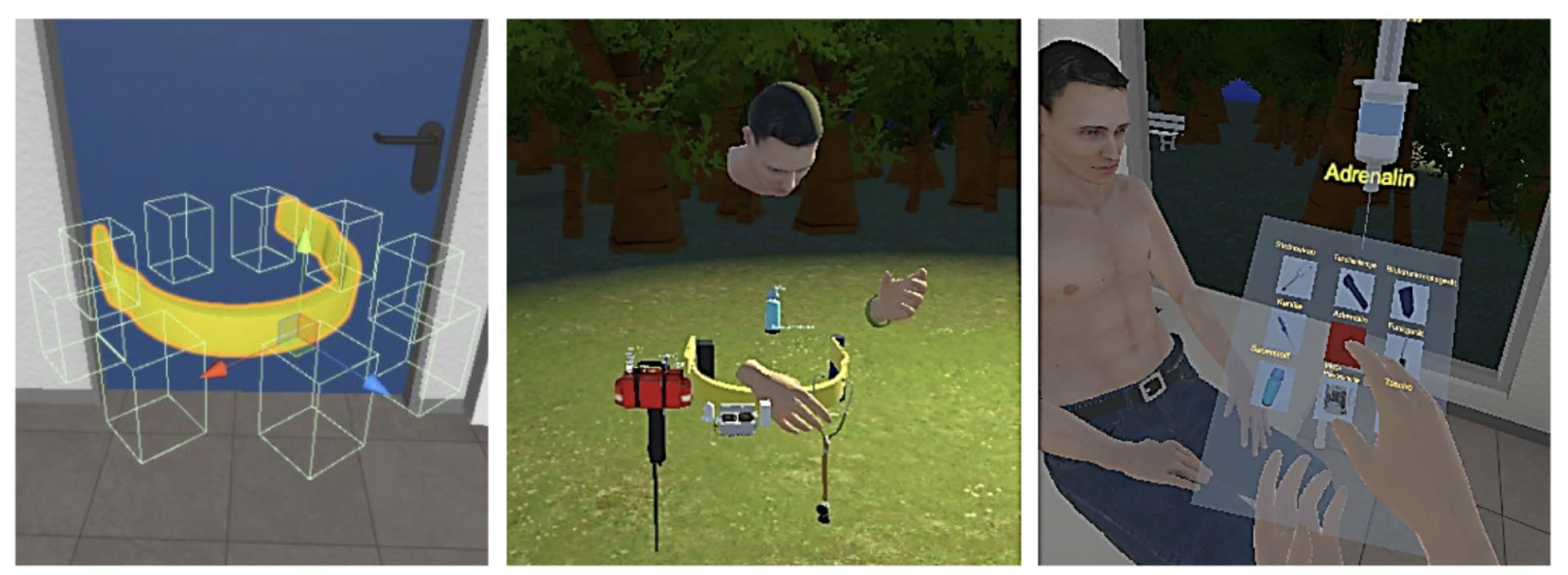}
  \caption{\centering Metaphorical inventory (belt) (left and middle) and Abstract inventory (hand table) (right) \cite{wegner2017comparison}.} 
  \label{fig.10: Metaphorical and Abstract inventory}
\end{figure}

Inventory is a term used to describe the possibilities of showing various items available in the VRE to its users during their user experience. Wegner et al. (2017) suggested two types of inventory concepts, the metaphorical and the abstract. In their prototype of metaphorical inventory, a virtual belt appears that consists of slots filled with items and rotated around the player’s waist. In their prototype of abstract inventory, a hand-held inventory menu appears where the player can switch on and off and can see at a glance which items are available for use. Fig. \ref{fig.10: Metaphorical and Abstract inventory}. 

They explored these two types in an experiment using a collaborative VRE game for paramedic vocational training. The sample was 24 paramedic trainees, in a between-groups mode, of ages between 20 and 33. Two players, acting as paramedics, must work together to treat a patient successfully, one plays an instructor and the other treats the patient using the available tools; the instructor communicates by reading instructions to the treating player through a tablet; the treating player cannot read the instructions on the tablet but has access to all tools and items in the game-play. The result showed no statistical significance in the preferences for abstract or metaphorical inventories which may be attributed to the lack of VR experience among the participants among other reasons. Based on their study, it should not be taken as a given that one type has the advantage of being favoured over the other \cite{wegner2017comparison}.

\subsection{Time and Spatial synchronization}\label{subsec5.2}

Since visual effects (including avatars and their positions in the virtual space) and audio effects are essential parts of a virtual reality user experience, their synchronization is a critical feature of a VRE.  It takes two forms, i.e., time synchronization and spatial synchronization of audio and video \cite{zhao2017study}. In the case of time synchronization, the term suggests that even 0.1 second of delay of one over the other has a serious negative effect in the user’s experience \cite{landy2010}. As an analogy, one may think of the troublesome experience of watching a movie in which the audio is trailing the video by 0.1 second. In the case of spatial synchronization, the term suggests that it is not enough for the audio and the video aspects of the VRE to be fully synchronized, as they must also match the positions of each other. For example, the audio should change its intensity as the user’s avatar is moving closer to or away from a particular object emanating that audio \cite{zhao2017study}.

In order to secure sufficient spatial synchronization of audio and video there are some specification requirements to meet for each. One such requirement is for the bitrate of the panoramic video to be high enough for the users to feel a high level of  immersion \cite{landy2010}.

Another key point here is that designing high fidelity avatars helps improve synchronization and behavioural realism, although very precise synchronization is not necessary. The problem is that low fidelity avatars, i.e., avatars with low quality in detail and e.g. lack of moving body parts, decrease the sense of co-presence, causing a feeling of loneliness undermining the quality of the VR experience \cite{hoppe2021shisha}. However, it must be noted, high fidelity avatars cannot substitute the benefits of quality collaboration amongst the VR users \cite{sousa2019negative}.

\subsection{Software, Hardware, and Connectivity}\label{subsec5.3}

One of the major challenges, in particular in the case of symmetric collaboration in a VRE, is to establish and suggest the optimum number of parallel users, either co-located or through distant collaboration, and the minimum requirements of software, hardware, and, especially, internet connectivity. Given the highly demanding nature of a typical VRE, it is highly imperative that the users’ internet connection allows for low latency network connectivity, i.e., the time it takes to send and receive a message between two users should be short, and usually needs to be below 15ms, ideally in the range of 4-7 ms \cite{elvezio2018collaborative}. These specifications affect the quality of the VRE. The users share the same VR experience, however, due to different internet connectivity and computer or application settings, their level of satisfaction may be considerably different.

\section{Conclusion and Future Work}\label{sec6}

Wang et al. (2021)  have identified three levels of authenticity that must be all addressed when creating a VR experience, namely, the authenticity of narrative, the authenticity of environment, and the authenticity of action \cite{wang2021authenticity}. The authenticity of narrative is easy to tackle, as this term refers to the scenario(s) that the VR experience is attempting to suggest and is also tied to the learning goals it attempts to achieve. It is not related to any of the technical aspects of VR, but merely about promoting the motivation and interest of the users involved. 

The authenticity of environment is the attempt to make sure that the environment resembles as closely as possible the actual environment and concepts in the form of 3D objects that the users must experience. Most studies, with some exceptions, focus on whether the users are more satisfied with the VR experience or with the traditional forms of human activity including learning; of course, the results are always positive towards the VR experience. The problem is there are rarely studies assessing the authenticity of action, i.e., the level of satisfaction of the VR users with the various actions they are allowed to take during their VR experience perhaps in collaboration with others. 

As discussed in this study, there are a number of different aspects of a VR experience that can be considered as variables/constructs in possible experiments. They belong to the general categories of spatial design, collaboration, visual and audio fidelity. This state-of-the-art study suggested a number of them that could be further studied as to their effect in a VRE. (The table in Appendix \ref{Appendix} outlines some experiments from a survey of the literature review). The following is a non-exhaustive list of those described in this study:

\begin{itemize}
\item Mutual embodiment/Shared perspective: The effect of synchronizing users’ actions and share attention of the 3D objects in the VRE.
\item Verbal communication: The value of the feature of verbal communication to the users during their VR experience.
\item Spatial details: The preferred level of details of the VRE both in its 3D objects and its avatars.
\item Teleportation as an option: The contribution of the feature of teleportation during the VR experience.
\item Teleportation technical requirements: The effect of the 4 attributes of teleportation during the users’ VR experience.
\item Teleportation visualization: The value of each of the 4 types of visualization of teleportation or combination of them.
\item Teleportation and scaling of 3D objects: The contribution of the dynamic scaling of 3D objects during teleportation.
\item Group teleportation: The preferred method of group teleportation.
\item Haptic feedback during users’ interaction: The effect of haptic feedback (vibrations) in the user’s experience when interacting with the VRE.
\item Sound feedback during users’ interaction: The value of adding sound feedback during the interaction of a user with the VRE.
\item Take-over control: Evaluation of the various methods of take-over control between the VR users.
\item Symmetric v. Asymmetric collaboration: The preferred choice of collaboration between two VR users.
\item Physical co-location v. distant collaboration: The preferred form of collaboration in a VRE in terms of the physical users.
\item Forms of co-habitation in a physical co-location: The preferred form of co-habitation method of multiple users following the physical co-location environment.
\item Inventory: The preferred type of inventory in a VRE.
\item Time synchronization of audio and video: The accepted level of time synchronization between audio and video during a VR experience.
\item Spatial synchronization of audio and video: The accepted level of spatial synchronization between audio and video during a VR experience.
\end{itemize}

There are two different ways to measure the level of satisfaction of users of the VR experience, namely, their engagement during the experience, or their post-experiment feedback evaluation, usually in the form of quantitative surveys, or both. Since the constructs are numerous and it is necessary to have a particular number of participants for the study of each of the constructs, it is very difficult, if not impractical, to plan for research that will combine more than 2 or 3 of the above constructs. Also, given that the application of VR is offered in a large variety of settings, e.g., businesses, government organizations, health, education, etc., and in a variety of fields, thus, it is necessary to select the setting and the field for the research of the selected constructs.

The authors propose and, indeed, are planning to focus on the effect of haptic feedback and the value of symmetric and asymmetric collaboration in higher education in the fields of computer science and aviation engineering.

\section*{Declarations}\label{sec7}

\subsection*{Ethical Approval}
This is a state-of-art research paper and as such it does not involve any human or animal participation. Hence, there are no ethical approvals necessary.

\subsection*{Competing interests}
There are no conflicts of interest of any sort between any of the authors and the journal or other financial entities.

\subsection*{Author's contributions}
Ourania K. Xanthidou is the main author of this State-Of-Art paper. Her contribution involves all aspects of the paper. 
Dr. Nadine Aburumman contributed with parts of the text, provided many of the sources for the paper, guidance on writing the paper, proofreading and corrections, and strong feedback and insights on improvements towards the final version. 
Prof. Hanene Ben-Abdallah contributed with some small but necessary pieces of the text, guidance on writing the paper and feedback and insights on improvements towards the final version.

\subsection*{Funding}
The authors have not received any kind of funding for this paper from any source.

\section*{Appendix}\label{Appendix}
\begin{longtblr}
[
  caption = {Survey of experiments on the VRE (2011-2023)},
  label = {tab:table},
]
{
  width = \linewidth,
  colspec = {XXX[1]},
  rowhead = 1,
  hlines,
  row{even} = {white},
  row{1} = {gray9},
  cells = {t},
  cell{1}{1} = {c},
  cell{1}{2} = {c},
  cell{1}{3} = {c},
} 

    \textbf{Reference/Aim/ Objectives} &
    \textbf{Experiment/Sample/ Methods} &
    \textbf{Findings}\\
Ricci et al. (2023) \newline
How immersive VR (IVR) and desktop VR (DVR) affect the shopping experience in the context of:\newline
1.	The duration of shopping experience.\newline
2.	Hedonic and utilitarian experiences.\newline
3.	The cognitive load.\newline
4.	The user experience.
& 
1.	Explore a visual fashion shop and interact with it.\newline
2.	Unity, Oculus Quest 2.\newline
3.	60 academics/students, within-subjects, ages 22-58.\newline
4.	Shapiro-Wilk test, Paired sample T-test, Mann-Whitney U test.
& 
1.	IVR causes longer duration regardless of prior VR experience.\newline
2.	2D experience is insufficient for delivering hedonic experience.\newline
3.	Cognitive load is comparable in both cases.\newline
4.	Indication of correlation between immersiveness and hedonic and utilitarian experiences.
\\
Fidalgo et al. (2023)\newline
Evaluation of MAGIC in terms of  face-to-face pointing gestures, Higher Pointing Agreement (HPA) and Perceived Message Understanding (PMU) in remote collaboration VRE. 
& 
1.	Running an evaluating MAGIC VRE.\newline
2.	Unity 3D, Oculus Rift CV1.\newline
3.	12 participants in pairs, within-subjects, physically co-located, \newline
4.	Shapiro-Wilk test, Pair T-test, Wilcoxon signed test.
& 
1.	Better understanding of the collaboration tasks in the case of HPA.\newline
2.	No difference between HPA and PMU, and in completion time. \newline
3.	Improved understanding when the user experienced the VRE under MAGIC.
\\
Wikstrom et al.  (2022)\newline
Evaluation of CoBlok puzzle: 2 users receive cards with different 2D views and identify the correct 3D shape with no other tools, only verbal communication in the VRE.
& - 
& 
Faster bandwidth and decreased latency, better cameras and microphones improve the collaborative VR experience.
\\
Drey et al. (2022)\newline
Symmetric vs. asymmetric collaborative VR experience: \newline
1.	Their impact on co-presence, immersion, player experience, motivation, cognitive load, and learning outcome.\newline
2.	Their impact of signaling on co-presence, immersion, player experience, motivation, cognitive load, and learning outcome. 
& 
1.	Student performs all action, teacher joins with a tablet asymmetrically. \newline
2.	Signaling guides attention of the student and is on/off based on learning progress.\newline
3.	Unity, HTC Vive Pro, tablet MS Surface 2 Pro, Dark Rift Networking 2.\newline
4.	46 subjects, between-subjects, ages 20-29, convenience sampling. 
& 
1.	Learning more pleasant and immersive for students/teachers when the teacher in symmetric.\newline
2.	Symmetric more motivating for the students; asymmetric equally satisfying learning outcomes.\newline
3.	Teachers in asymmetric rely on students’ orientation which cost miscommunication.\newline
4.	Signaling affect students more than teachers as the latter are not actively engaged.
\\

Hoppe et al. (2021)\newline
Comparison between ShiSha and Vishnu VREs in the context of:\newline
1.	Higher usability.\newline
2.	Higher task performance.\newline
3.	Higher co-presence.
& 
1.	ShiSha vs Vishnu VRE puzzles.\newline
2.	Unity, SteamVR, HTC Vive Pro, Windows 10. \newline
3.	32 participants.\newline
4.	ANOVA, Wilcoxon signed-rank test, Pearson’s correlation.
 
& 
1.	Shisha has higher usability.\newline
2.	Shisha has higher task performance.\newline
3.	Shisha has higher co-presence.

\\
Weissker et al. (2020)\newline
A framework for group navigation that includes navigational groups (forming), distribute navigational responsibilities (norming), navigate together (performing), and eventually split up again (adjourning).
& 
1.	32 tasks, 1 per pair side-by-side or in a queue.\newline
2.	40 students/employees, ages 20-38.\newline
3.	HTC Vive Pro.\newline
4.	Shapiro-Wilk and QQ-Plots test.
& 
1.	Task completion time in the Adjust condition better than the Baseline.\newline
2.	Physical walking distances smaller in the Adjust condition.\newline
3.	Task load score smaller in the Adjust condition.\newline
4.	Group navigation helps the users focus, discuss, evaluate, and resolve problems.

\\
deBack et al. (2020)\newline
Examine the learning benefits using the CAVE mode compared to traditional textbook in 2-question tests with 20 four-option MCQs in a VRE.
 
& 
1.	Unity.\newline
2.	40 participants, groups of 2-4, between-subjects, ages 18-67.\newline
3.	One-way ANOVA, Kolmogorov-Smirnov.

& 
CAVE VRE results in better learning than traditional textbook.
\\
Grandi et al. (2019)\newline
An assessment of asymmetric interaction in collaborative VRE through a 3D docking task to align the position, rotation, and scale of virtual objects. 
& 
1.	Unity, UNET API.\newline
2.	36 participants, within-subjects.\newline
3.	ANOVA, Wilcoxon signed-rank test, Shapiro-Wilk.
& 
1.	No increase in speed during the task when working in pairs.\newline
2.	Work in pairs reduces the workload.\newline
3.	VR-VR performs better than AR-AR and VR-AR.\newline
\\
Ardal et al. (2019)\newline
A tool that provides immersive previsualizations (previs) for filmmakers to create, discuss and edit scenes, in a collaborative or in a solo VRE.
& 
1.	Unity, HTC Vive Pro.\newline
2.	20 filmmakers, within-subjects, ages 18-59.\newline
3.	Repeated measures ANOVA for the questionnaire, Chi-Square for the comparison between solo and collaborative setup, Bivariate correlation (Pearson’s correlation) for the personality traits.
& 
1.	Collaborative VR can be successful.\newline
2.	Collaborative VRE is more fun and preferred.\newline
3.	Solo VRE showed improved concentration levels and performance.\newline
4.	Collaborative VRE successful with participants as avatars with audio feedback.\newline
5.	Splitting roles and interfaces benefits the collaboration aspects.
\\
Xia et al. (2018)\newline
Introduce and evaluate the parallel objects technique of SPACETIME, an editing tool for collaborative VRE; ability to translate, scale, or rotate 3D objects.
 
& 
1.	Unity, Oculus VR SDK, Photon Unity.\newline
2.	Teleportation stored in a stack allows return to previous positions. \newline
3.	6 subjects, ages 22-41.

& 
The technique of parallel objects and the concept of a container is beneficial for the collaboration in a VRE.
\\
Rios et al. (2018)\newline
Study the user’s locomotor behavior in a collaborative VRE.
  
& 
1.	A puzzle game in which the users walk in 2x4m indoor lab, grab pieces of the puzzle and place them on the board.\newline
2.	Unity.\newline
3.	8 participants in pairs, ages 20-40.\newline
4.	ANOVA for the questionnaire evaluation.
 
& 
1. Synchronized Animations and footsteps sound do not affect performance.\newline
2. The users in VREs more cautious than in reality, regardless of how realistic and accurate the VRE looks as they worried about colliding with others.\newline
3.	An indication of different intensity of caution in the VRE even with small anomalies in the synchronization of animation and sound.
\\
Liu and Kaplan (2018)\newline
Study of situation awareness and conversational grounding impacted by the loss of visual information in a collaborative VRE of Minesweeper game in 5 modes: Desktop Freeplay (DF), VR Freeplay (VRF), Competitive VRE within Browser (CVREB), VR Split (VRS), and VR Turn-Based (VRTB).
 
& 
1. The 5 modes are: \newline
a.	DF: play using keyboard, mouse, and monitor.\newline
b.	VRF: play through the browser within the VRE.\newline
c.	CVREB: play through the browser within the VRE, see other players’ biometrics.\newline
d.	VRS: full collaboration, cannot see opponents.\newline
e.	VRTB: full collaboration, can see opponents biometric and progress. \newline
2.	Samsung GearVR, Oculus Rift.\newline
3.	16 students in pairs, ages 20-35.
& 
1.	Physically co-located subjects move and talk less when playing in solo.\newline
2.	Players get more stressed when seeing others’ players avatars or biometrics.\newline
3.	Collaborating in the VRE fosters more communication and leads to less movements.\newline
4.	Solo participation leads to lower sense of urgency.
\\
Wegner et al. (2017)\newline
Explore metaphorical and abstract inventory concepts through a collaborative VRE for paramedic vocational training in which One user reading instructions from a tablet and the other treating a patient in the VRE.
 
& 
1.	Abstract inventory as table menu that can be switch on/off. \newline
2.	Metaphorical inventory as virtual belt around the waist with slots of available items.\newline
3.	24 subjects, between-groups, ages 20-33.\newline
4.	HTC VIVE, Independent samples t-test.
& 
There is no preference between abstract or metaphorical inventories.
\\

Schwind et al. (2017)\newline
Investigate the effect of different styles of virtual hands (VH), i.e., human and non-human, and gender on presence experienced by men and women in immersive VRE.
  
& 
1.	3 different tasks in VR experience.\newline
2.	VR questionnaire appears within the VRE.\newline
3.	Unity, Oculus Rift DK2, and leap motion sensor.\newline
4.	28 students and employees.
& 
1. Participants feel uncomfortable using non-human VH. \newline
2. Females feel uncomfortable with male VH; males using female VH find it realistic and unusual but attractive. \newline
3. Participants feel uncomfortable with VH movement not synchronized.\newline
4. Female users feel higher levels of co-presence using non-human VH.
\\
Lacoche and Pallamin (2017)\newline
Address the cohabitation and safety problem in a shared physical space when collaborating in a VRE shooter game experiment in different modes for the separation of avatars.
  
& 
1. 4 modes of separation:  Extended Grid (EG), Ghost Avatar (GA), Navigation Floor (SNF), Separated Tracked Spaces (STS).\newline
2.	Unity, SteamVR, Photon, HTC Vive, MSI VR One.\newline
3.	38 subjects in pairs, ages 14-59.\newline
4.	ANOVA, Post hoc t-test.
& 
1.	Users’ safety is not better with STS because they are forced to keep looking at the floor instead of the game or each other.\newline
2.	Users do not feel more free to move using EG, GA, and SNF compared to STS.\newline
3.	Users prefer EG, GA, SNF than being constrained in an STS.
\\
Moulec et al. (2016)\newline
Test the smooth compared to the sudden Take-Over control mechanism in a Collaborative Virtual Environments for Training (CVET) maritime navigation VRE.

& 
1.	A desktop collaborative VRE, using joystick, for maritime navigation training.\newline
2.	8 participants in pairs, within-subjects.\newline
3.	Non-parametric Wilcoxon pairwise test.

& 
1.	Participants prefer smooth over the sudden take-over control.\newline
2.	No difference in the easiness.\newline
3.	Participants feel increased awareness and information transfer when visual feedback is available.
\\
Hrimech (2011)\newline
Study the effect of 3D interaction metaphors on the user experience in a VRE 3D puzzle.
  
& 
1.	No visual contact between users, only verbal communication.\newline
2.	3 interaction metaphors: \newline
a.	Virtual Hand: represents physical hand, follows physical movements.\newline
b.	Ray Casting: allows selection of a 3D object pointing it with a virtual ray.\newline
c.	GoGo: an elastic virtual arm with which users can interact with objects in distance.\newline
3.	32 participants in pairs, ages 18-57.\newline
4.	One-Way ANOVA, Pearson’s correlation.
 
& 
1.	Ray-Casting increases co-presence and awareness compared to the other metaphors.\newline
2.	Users have a feeling of disappointment with the lack of animation with the virtual hand metaphor.\newline
3.	Ray-Casting metaphor leads to higher involvement than the virtual hand.\newline
4.	Collaborative effort is more intense with the GoGo metaphor.
\\

\end{longtblr}

\bibliographystyle{unsrt}  
\bibliography{references}

\end{document}